\documentclass[aps,prl,twocolumn,showpacs,superscriptaddress]{revtex4}
\usepackage{graphicx}

\begin{document}

\title{Antiferromagnetic spin excitations in single crystals of
nonsuperconducting Li$_{1-x}$FeAs}
\author{Meng Wang}
\affiliation{Institute of Physics, Chinese Academy of Sciences, P. O. Box 603, Beijing
100190, China }
\affiliation{Department of Physics and Astronomy, The University of Tennessee, Knoxville,
Tennessee 37996-1200, USA }
\author{X. C. Wang}
\affiliation{Institute of Physics, Chinese Academy of Sciences, P. O. Box 603, Beijing
100190, China }
\author{D. L. Abernathy}
\affiliation{Neutron Scattering Science Division, Oak Ridge National Laboratory, Oak
Ridge, Tennessee 37831-6393, USA}
\author{L. W. Harriger}
\affiliation{Department of Physics and Astronomy, The University of Tennessee, Knoxville,
Tennessee 37996-1200, USA }
\author{H. Q. Luo}
\affiliation{Institute of Physics, Chinese Academy of Sciences, P. O. Box 603, Beijing
100190, China }
\author{Yang Zhao}
\affiliation{NIST Center for Neutron Research, National Institute of Standards and
Technology, Gaithersburg, MD 20899, USA}
\affiliation{
Department of Materials Science and Engineering, University of Maryland, College Park, MD, 20742, USA
}
\author{J. W. Lynn}
\affiliation{NIST Center for Neutron Research, National Institute of Standards and
Technology, Gaithersburg, MD 20899, USA}
\author{Q. Q. Liu}
\affiliation{Institute of Physics, Chinese Academy of Sciences, P. O. Box 603, Beijing
100190, China }
\author{C. Q. Jin}
\affiliation{Institute of Physics, Chinese Academy of Sciences, P. O. Box 603, Beijing
100190, China }
\author{Chen Fang}
\affiliation{Department of Physics, Purdue University, West Lafayette, Indiana 47907, USA}
\author{Jiangping Hu}
\affiliation{Department of Physics, Purdue University, West Lafayette, Indiana 47907, USA}
\affiliation{Institute of Physics, Chinese Academy of Sciences, P. O. Box 603, Beijing
100190, China }
\author{Pengcheng Dai}
\email{pdai@utk.edu}
\affiliation{Department of Physics and Astronomy, The University of Tennessee, Knoxville,
Tennessee 37996-1200, USA }
\affiliation{Neutron Scattering Science Division, Oak Ridge National Laboratory, Oak
Ridge, Tennessee 37831-6393, USA}
\affiliation{Institute of Physics, Chinese Academy of Sciences, P. O. Box 603, Beijing
100190, China }

\begin{abstract}
We use neutron scattering to determine spin excitations in single crystals
of nonsuperconducting Li$_{1-x}$FeAs throughout the Brillouin zone.
Although angle resolved photoemission experiments and local density
approximation calculations suggest poor Fermi surface nesting conditions for
antiferromagnetic (AF) order, spin excitations in Li$_{1-x}$FeAs occur at
the AF wave vectors $Q=(1,0)$ at low energies, but move to wave
vectors $Q=(\pm 0.5,\pm 0.5)$ near the zone boundary with a total magnetic
bandwidth comparable to that of BaFe$_{2}$As$_{2}$. These results reveal
that AF spin excitations still dominate the low-energy physics of these
materials and suggest both itinerancy and strong electron-electron
correlations are essential to understand the measured magnetic excitations.
\end{abstract}

\maketitle




Understanding whether magnetism is responsible for superconductivity in
FeAs-based materials continues to be one of the most important unresolved
problems in modern condensed matter physics \cite{johnston,lumsden,lynn}.
For a typical iron arsenide such as LaFeAsO \cite{kamihara}, band structure
calculations predict the presence of the hole-like Fermi surfaces at the 
$\Gamma (0,0)$ point and electron-like Fermi surfaces at the $M(1,0)/(0,1)$
points in the Brillouin zone [Fig. 1(a)] \cite{mazin}. As a consequence,
Fermi surface nesting and quasiparticle excitations between the hole and
electron pockets can give rise to static antiferromagnetic (AF)
spin-density-wave order at the in-plane wave vector ${Q}=(1,0)$ 
\cite{jdong}. Indeed, neutron scattering experiments have shown the presence of
the ${Q}=(1,0)$ AF order in the parent compounds of iron arsenide
superconductors, and doping to induce superconductivity suppresses the
static AF order \cite{cruz}. In addition, angle resolved photoemission
measurements \cite{hding} have confirmed the expected hole and
electron pockets in superconducting iron arsenides, thus providing evidence
for superconductivity arising from the sign revised electron-hole
inter-pocket quasiparticle excitations \cite{mazin,seo2008,kuroki,chubkov,fwang}.

Of all the FeAs-based superconductors \cite{johnston}, LiFeAs is special
since it has the highest transition temperature ($T_{c}=18$ K) amongst the
stoichiometric compounds \cite{xcwang,jhtapp,mjpitcher,flpratt,cwchu}.
Furthermore, it does not have static AF order due to the poor Fermi surface
nesting properties with shallow hole pockets near the $\Gamma (0,0)$ 
\cite{borisenko}. It has been suggested that the flat tops of the hole pockets in
LiFeAs imply a large density of states near the Fermi surface, which should
promote ferromagnetic (FM), instead of the usual AF, spin fluctuations for
superconductivity \cite{brydon}. If this is indeed the case, AF spin
fluctuations should not be fundamental to the superconductivity of
FeAs-based materials and the superconducting pairing would not be in the
spin singlet channel. A determination of the magnetic properties in LiFeAs
is thus important to complete our understanding about the role of magnetism
in the superconductivity of FeAs-based materials.

In this paper, we present inelastic neutron scattering measurements on
single crystals of nonsuperconducting Li$_{1-x}$FeAs with $x=0.06\pm 0.01$, where there is no
static AF order. As a function of increasing energy, spin excitations in Li$
_{0.94}$FeAs have a spin gap below $\Delta \approx 13$ meV, are centered at
the AF wave vector $Q=(1,0)$ for energies up to $\sim $80 meV, and then
split into two vertical bands of scattering before moving into the zone
boundaries at the wave vectors $Q^{\prime }=(\pm 1/2,\pm 1/2)$ 
near $E\approx 130$ meV. These $Q^{\prime }$ vectors have been observed in the
spin excitations of FeTe/Se compounds and imply the existence of strong
competition between FM and AF exchange couplings \cite{Lip2011}. While the
dispersions of the low-energy spin excitations ($E\leq 80$ meV) in Li
$_{0.94} $FeAs are similar to that of (Ba,Ca,Sr)Fe$_{2}$As$_{2}$ 
\cite{harriger,jzhao,ewings}, the high-energy spin excitations near the
zone boundary are quite different from these materials, and cannot be
modeled from a simple Heisenberg Hamiltonian with effective nearest ($J_{1a}$ and $J_{1b}$) and next
nearest neighbor ($J_2$) exchange couplings \cite{harriger,jzhao}. By integrating
the local susceptibility $\chi ^{\prime \prime }(\omega )$ in absolute units
over the entire bandwidth of spin excitations, we find the spin fluctuating
moment $\left\langle m^{2}\right\rangle =2.1\pm 0.6\ \mu _{B}^{2}$, a value
that is comparable with other pnictides. Therefore, spin excitations in 
Li$_{0.94}$FeAs are similar to other iron pnictides but are not directly
associated with Fermi surface nesting from hole and electron pockets,
contrary to expectations from local density approximation calculations \cite{borisenko,brydon}.

\begin{figure}[t]
\includegraphics[scale=0.4]{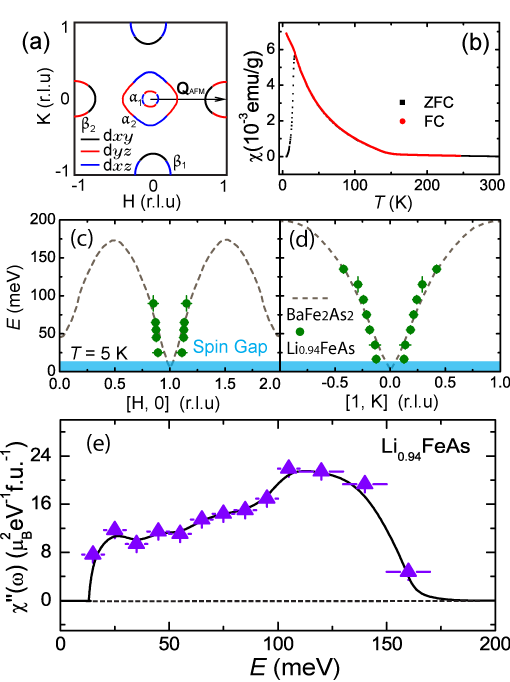}
\caption{ (color online). (a) Fermi surfaces from the
spin-restricted local density approximation calculation for LiFeAs 
\protect\cite{borisenko,brydon}. There are two hole-like Fermi surfaces near the 
$\Gamma (0,0)$ point with $d_{yz}$/$d_{xz}$ character and one electron-like
Fermi surface near the $M(1,0)$ point. The nesting condition for the
expected AF nesting wave vector $Q_{AFM}=(1,0)$ is not favorable 
\protect\cite{borisenko}. (b) Zero field cooled (ZFC) and field cooled (FC)
susceptibility measurements on Li$_{0.94}$FeAs. No superconductivity was
observed due to Li deficiency. (c) The dashed lines show spin wave
dispersions along the $[H,0]$ and $[1,K]$ directions for BaFe$_{2}$As$_{2}$
at 5 K \protect\cite{harriger}. The filled circles show the measured spin
excitation dispersions along the $[H,0]$ and $[1,K]$ directions for 
Li$_{0.94}$FeAs.  While spin waves in BaFe$_{2}$As$_{2}$ extend up to 200 meV
along the $[1,K]$ direction, spin excitations in Li$_{0.94}$FeAs reach the
zone boundary near $Q=(1,0.5)$. }
\end{figure}

Our experiments were carried out on the ARCS time-of-flight chopper
spectrometer at the spallation neutron source, Oak Ridge National
laboratory. We also performed thermal triple-axis spectrometer measurements
on the BT-7 triple-axis spectrometer at NIST center for neutron Research.
Our single crystals were grown using the flux method and
inductively coupled plasma analysis on the samples showed that the
compositions of the crystals are Li$_{0.94\pm 0.01}$FeAs. Figure 1(b) shows
zero field cooled (ZFC) and field cooled (FC) susceptibility measurements on
Li$_{0.94}$FeAs, which indicate spin glass behavior with no evidence for
superconductivity. To study the spin excitations, we co-aligned 7.5 g of
single crystals of Li$_{0.94}$FeAs (with a mosaic of 2$^{\circ }$) and
loaded the samples inside a He refrigerator or cryostat. To facilitate easy
comparison with spin wave measurements in BaFe$_{2}$As$_{2}$ \cite{harriger}%
, we defined the wave vector $Q$ at ($q_{x}$, $q_{y}$, $q_{z}$) as $%
(H,K,L)=(q_{x}a/2\pi ,q_{y}b/2\pi ,q_{z}c/2\pi )$ reciprocal lattice units
(rlu), where $a=b=5.316$ \AA , and $c=6.306$ \AA . For both triple-axis and
ARCS measurements, we aligned crystals in the $[H,0,L]$ scattering zone. The
ARCS data are normalized to absolute units using a vanadium standard. The
incident beam energies were $E_{i}=80,140,250$ meV with $E_{i}$ parallel to
the $c$-axis.

\begin{figure}[t]
\includegraphics[scale=0.4]{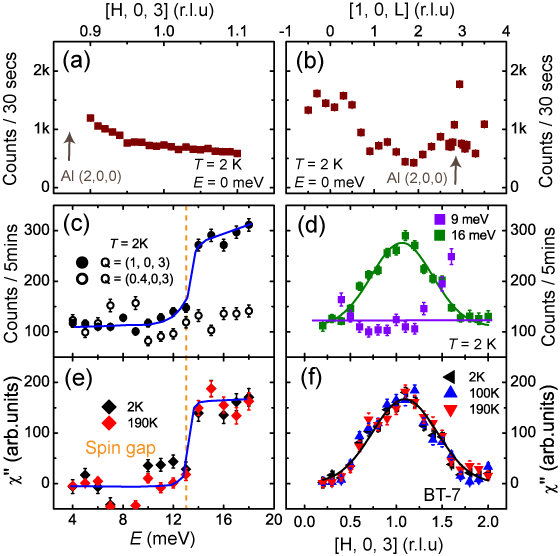}
\caption{ (color online). Triple-axis measurements to search for static AF
order and spin excitations in Li$_{0.94}$FeAs. (a) Elastic scattering along
the $[H,0,3]$ and (b) $[1,0,L]$ directions at 2 K show no evidence of
AF order at the expected position $Q=(1,0,3)$. The arrows indicate Al
sample holder scattering. (c) Constant-$Q$ scans at the wave vectors 
$Q=(1,0,3)$ (signal) and $Q=(0.4,0,3)$ (background) positions at 2 K.  A clear
spin gap is seen at $\Delta =13$ meV. (d) Constant-energy scans at $E=9$, 16
meV along the $[H,0,3]$ direction. While the scan at $E=9$ meV is
featureless, a clear peak is seen at $E=16$ meV confirming the spin gap. (e)
Imaginary part of the dynamic susceptibility $\protect\chi ^{\prime \prime }$
at 2 K and 190 K. The magnitude of the spin gap is unchanged between 2 and
190 K. (f) Temperature dependence of the $\protect\chi ^{\prime \prime }(Q)$
at $E=16$ meV for 2, 100, and 190 K. The $\protect\chi ^{\prime \prime }(Q)$
is almost temperature independent between 2 K and 190 K. Error bars where
indicated represent one standard deviation.}
\end{figure}

Before describing in detail the spin excitation dispersion curves and
dynamic local susceptibility in Figs. 1 (c)-1(e), we first discuss the
triple-axis measurements on the static AF order and spin excitations.
Figures 1(a) and 1(b) show elastic scattering along the $[H,0,3]$ and $%
[1,0,L]$ directions across the expected AF peak position $(1,0,3)$,
respectively. In contrast to Na$_{1-x}$FeAs, where static AF order is
clearly observed \cite{slli}, there is no evidence for static AF order in
this sample. To search for AF spin excitations, we carried out constant-$Q$
scans at the AF wave vector ${Q}=(1,0,3)$ and background $(0.4,0,3)$
positions. The outcome in Fig. 2(c) shows a step-like increase in scattering
above background for $E>13$ meV, clearly suggesting the presence of a large
spin gap of $\Delta =13$ meV. To confirm there is indeed a spin gap, we
carried out constant-energy scans along the $[H,0,3]$ direction at $E=9$ and
16 meV as shown in Fig. 1(d). While the scattering is featureless at $E=9$
meV, there is a clear peak centered at ${Q}=(1,0,3)$ at $E=16$ meV.
Figure 3(e) shows temperature dependence of the imaginary part of dynamic
susceptibility $\chi ^{\prime \prime }(E)$ obtained by subtracting the
background and correcting for the Bose population
factor. Surprisingly, the spin gap has no observable temperature dependence
between 2 K and 190 K, much different from the temperature dependence of the
spin gaps in the (Ba,Sr,Ca)Fe$_{2}$As$_{2}$ family of materials 
\cite{matan,zhaoprl,rob}, which disappear rapidly with increasing temperature.
The weak temperature dependence of the dynamic susceptibility has been
confirmed by constant-energy scans in Fig. 2(f), where $\chi ^{\prime \prime
}(Q)$ at $E=16$ meV remains essentially unchanged from 2 K to 190 K.

\begin{figure}[t]
\includegraphics[scale=0.45]{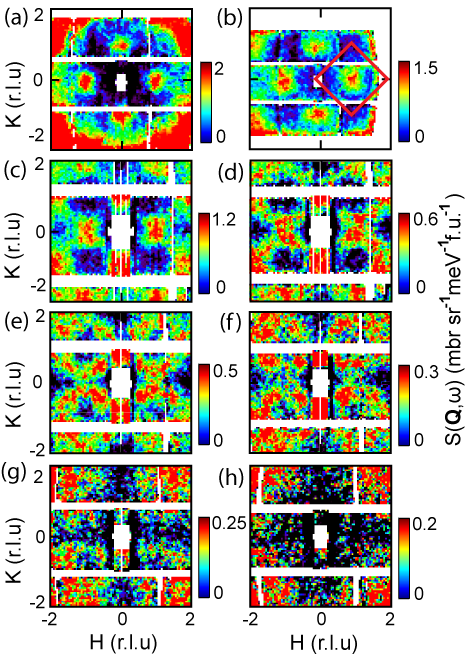}
\caption{(color online). Constant-energy images of the scattering in the $%
[H,K]$ zone as a function of increasing energy for Li$_{0.94}$FeAs at energy
transfers of (a) $E=25\pm 5$ meV; (b) $45\pm 5$ meV (with $E_{i}=80$ meV);
(c) $70\pm 10$ mV; (d) $90\pm 10$ meV; (e) $110\pm 10$ meV; (f) $130\pm 10$
meV; (g) $150\pm 10$ meV; (h) $170\pm 10$ meV, all with $E_{i}=250$ meV. The
scattering intensity is in absolute units. The box in (b) shows the
Brillouin zone used to integrated the susceptibility. }
\end{figure}

\begin{figure}[t]
\includegraphics[scale=0.4]{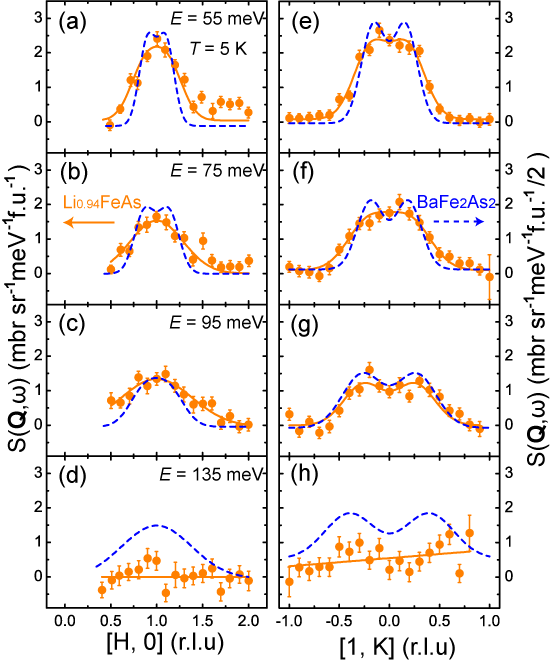}
\caption{(color online). Constant-energy cuts of the spin excitation
dispersion as a function of increasing energy along the $[H,0]$ and $[1,K]$ 
directions for Li$_{0.94}$FeAs. The dashed curves show identical cuts
for spin waves of BaFe$_{2}$As$_{2}$ divided by 2 \protect\cite{harriger}. Both are in
absolute units. Constant-energy cuts along the $[H,0]$ direction at (a) 
$55\pm 5$, (b) $75\pm 5$, (c) $95\pm 10$, and (d) $135\pm 10$ meV. Similar
cuts along the $[1,K,]$ direction are shown in (e)-(h). The dynamic spin
correlation length is $\protect\xi \approx 12\pm 3$ \AA. }
\end{figure}

Figure 3 summarizes the ARCS time-of-flight measurements on Li$_{0.94}$FeAs
at 5 K. Since spin excitations in Li$_{0.94}$FeAs have no $c$-axis
modulations, we show in Figs. 3(a)-3(h) two-dimensional constant-energy ($E$) 
images of the scattering in the $(H,K)$ plane for $E=25\pm 5$,
$45\pm 5$, $70\pm 10$, $90\pm 10$, $110\pm 10$, $130\pm 10$, 
$150\pm 10$, and $170\pm 10$ meV, respectively. For
energies between $25\pm 5 \leq E \leq 90\pm 10$ meV, spin excitations form
transversely elongated ellipses centered around AF ${Q} = (1,0)$. The intensity of spin excitations decrease with
increasing energy, this is remarkably similar to spin waves in BaFe$_2$As$_2$
\cite{harriger}. For energies above 90 meV, spin excitations split into two
horizontal arcs that separate further with increasing energy. The
excitations finally merge into ${Q} = (m\pm 0.5,n\pm
0.5)$ ($m,n=0,1,2$) and become weaker above 150 meV in Fig. 3(g) and 3(h).

In order to determine the dispersion of spin excitations for Li$_{0.94}$
FeAs, we show in Fig. 4 cuts through the two-dimensional images in Fig. 3
and compare with identical cuts
for BaFe$_{2}$As$_{2}$. \ Figures 4(a)-4(d) show constant-energy cuts along
the $[H,0]$ direction for energies of $E=55\pm 5$, $75\pm 5$, $95\pm 5$, 
$135\pm 5$ meV, respectively, while the dashed lines show identical spin
wave cuts for BaFe$_{2}$As$_{2}$ \cite{harriger}. Since both measurements
were taken in absolute units, we can see that spin excitations in Li$_{0.94}$
FeAs are similar to that of BaFe$_{2}$As$_{2}$ per Fe 
below 95 meV \cite{harriger}. Figures 4(e)-4(h) show constant-energy cuts
along the $[1,K]$ direction for identical energies as that of Figs.
4(a)-4(d). For energies above 95 meV, the strength of the spin excitations
in Li$_{0.94}$FeAs are rapidly suppressed compared to those of BaFe$_{2}$As$
_{2}$ and become very weak above $E=135$ meV. This can originate from an
absence of magnetic scattering, or that the scattering is very broad as
might occur when an itinerant electron system interacting with Stoner
excitations. This is different from spin waves in BaFe$_{2}$As$_{2}$,
which extend up to 250 meV. Based on these constant-energy cuts, we show in
Figs. 1(c) and 1(d) the comparison of spin excitation dispersions of 
Li$_{0.94}$FeAs (filled circles) with those of spin waves in BaFe$_{2}$As$_{2}$
(dashed lines). They are similar for energies between $50-95$ meV, while the
spin excitations in Li$_{0.94}$FeAs are broader below 50 meV.

We have attempted, but failed, to use a simple Heisenberg Hamiltonian with
the effective nearest and next nearest neighbor   
exchange couplings to fit the observed spin 
excitations spectra \cite{harriger,jzhao}. For all the possible combinations of the $J_{1a}$, $J_{1b}$, 
and $J_{2}$, the expected zone boundary spin excitations are quite
different from the observed spectra (see supplementary information) 
\cite{suppl}. If we include the next-next nearest neighbor exchange coupling 
$J_{3}$, the expected spectra near the zone boundary have some resemblance to
the data in Fig. 3 although the low-energy excitations would be
different (see supplementary information). This means that the effective
exchange couplings in Li$_{0.94}$FeAs are extremely long-ranged, a hallmark
that itinerant electrons are important for spin excitations in this
material. Since the data close to the band top along the $[1,K]$ direction
are higher in energy than along the $[H,0]$ direction, we need $J_{1b}<0$ to
recover this feature in a $J_{1a}$-$J_{1b}$-$J_{2}$-$J_{3}$ model (see
supplementary information). This means that effective exchange interactions
in Li$_{0.94}$FeAs may be similar to the (Ca,Sr,Ba)Fe$_{2}$As$_{2}$ iron
pnictides \cite{harriger,jzhao,ewings} in spite of their different zone
boundary spectra.

Finally, we show in Figure 1(e) the energy dependence of the local
susceptibility, defined as $\chi ^{\prime \prime }(E)=\int {\chi^{\prime
\prime }(\mathbf{q},E)d\mathbf{q}}/\int {d\mathbf{q}}$, where the average is
over the magnetic scattering signal $\chi ^{\prime \prime }(\mathbf{q},E)$
over the Brillouin zone [Fig. 3(b)] \cite{lester}. The corresponding fluctuating moment $
\left\langle m^{2}\right\rangle =2.1\pm 0.6\ \mu _{B}^{2}$ per formula unit.
We can use both pure local and itinerant spin models to sketch a basic
physical picture based on the moment value. If we assume a quantum local
spin model to describe the fluctuations, the moment value implies the spin
value is about one. If we take a pure itinerant model, our result would
suggest that at least three electrons per iron site occupy the states with
energies up to the magnetic bandwidth ($\sim 150$ meV) below the Fermi
energy. This suggests that the bandwidths of the electron bands near the
Fermi surface are extremely narrow. In other words, the band renormalization
factors are large and the electron-electron correlations must be strong.

In summary, we measured spin excitations in single crystals of 
Li$_{0.94}$FeAs. Similar to other iron pnictides, the low energy excitations are still
strongly AF \cite{platt}. However, comparing to other iron pnictides, they
have several distinct properties: (a) a larger spin gap, close to $13$ meV that is essentially 
temperature independent below 190 K;
(b) a comparable total magnetic bandwidth; (c) different wave vectors
at the zone boundary for high energy excitations. Moreover, the excitations
can not be described by magnetic models with only short range magnetic
exchange couplings. Our results suggest the AF spin fluctuations are
fundamental to the superconductivity of FeAs-based materials. FM
fluctuations exist in Li$_{0.94}$FeAs, but they only affect the high energy
spin excitations.

During the process of writing up this paper, we became aware of a related
work on powder samples of superconducting LiFeAs, where AF spin fluctuations 
have been reported \cite{taylor}.

The work in IOP is supported by CAS, the MOST of China and NSFC. This work
is also supported by the U.S. DOE BES No. DE-FG02-05ER46202, and by the U.S.
DOE, Division of Scientific User Facilities.





\def\avg#1{\langle#1\rangle}
\def\Re {\mbox{Re}}
\def\Im {\mbox{Im}}
\def\be{\begin{equation}}       \def\ee{\end{equation}}
\def\bea{\begin{eqnarray}}      \def\eea{\end{eqnarray}}
\def\half{\frac{1}{2}}
\def\PRA{Phys. Rev. A~}
\def\PRB{Phys. Rev. B~}
\def\PRD{Phys. Rev. D~}
\def\RMP{Rev. Mod. Phys. ~}
\def\PRL{Phys. Rev. Lett.~}

\clearpage

\maketitle{\textbf{Supplementary Information: Antiferromagnetic spin excitations in single crystals of
nonsuperconducting Li$_{0.94}$FeAs}}
\\[5mm] 
\maketitle{Meng Wang, X. C. Wang, D. L. Abernathy, L. W. Harriger, H. Q. Luo, Yang Zhao, J. W. Lynn, Q. Q. Liu, C. Q. Jin, Chen Fang, Jiangping Hu, Pengcheng Dai }
\\[5mm]
\author{Meng Wang}
\author{X. C. Wang}
\author{D. L. Abernathy}
\author{L. W. Harriger}
\author{H. Q. Luo}
\author{Yang Zhao}
\author{J. W. Lynn}
\author{Q. Q. Liu}
\author{C. Q. Jin}
\author{Chen Fang}
\author{Jiangping Hu}
\author{Pengcheng Dai}


In order to test if a simple $J_{1a}$-$J_{1b}$-$J_2$-$J_3$ Heisenberg Hamiltonian can reproduce the observed
spin excitations spectra in Fig. 3, we simulated the expected spin wave spectra using Heisenberg Hamiltonian \cite{jzhao}.
To facilitate direct comparison with the data in Fig. 3, we normalized the calculated intesity at 90 meV to be the same
as that in Fig. 3. Therefore, the calculated spectra can be directly compared with the observed spectra.
Figure SI5 shows spin wave calculations assuming $SJ_{1a}=23$ meV, $SJ_{1b}=-4$ meV,
$SJ_{2}=13$ meV, and $SJ_3=-5$ meV.  While the zone boundary spectra have
some similarity to the data, the spectra clearly disagree with the data around intermediate energies.
Figure SI6 shows similar calculation assuming $SJ_{1a}=SJ_{1b}=10$ meV,
$SJ_{2}=20$ meV, and $SJ_3=-8$ meV.  Figure SI7 plots calculations assuming
$SJ_{1a}=23$ meV, $SJ_{1b}=-4$ meV,
$SJ_{2}=13$ meV, and $SJ_3=0$ meV; and Figure SI8 show calculation with
$SJ_{1a}=SJ_{1b}=10$ meV,
$SJ_{2}=20$ meV, and $SJ_3=0$ meV.  The spin wave band tops are in Figs. SI5-SI8 are
150 meV, 150 meV, 110 meV, and 100 meV, respectively.
To obtain similar scattering pattern as observed near the band top by a Heisenberg Hamiltonian with effective exchange couplings,
the next-next nearest neighbor exchange coupling $J_3$ has to be included.

We note that spin excitations close to the band top in Fig1(e) clearly shows that zone boundary energy in the $[1,K]$ direction is higher in energy than that along the $[H,0]$ direction. In the following we show that in a $J_{1a}$-$J_{1b}$-$J_2$-$J_3$ model with antiferromagnetic $J_{1a}$ and $J_2$ and ferromagnetic $J_3$, we need and only need $J_{1b}<0$ (ferromagnetic) to recover this feature.

The $J_{1a}$-$J_{1b}$-$J_2$-$J_3$ model is analogous to the $J_{1a}$-$J_{1b}$-$J_2$ spin model used to describe CaFe$_2$As$_2$\cite{jzhao}. In the following we will consider a detwinned system, and one should bear in mind that one q-point $(q_x,q_y)$ in a twinned system corresponds to $(q_x,q_y)$ and $(q_y,q_x)$ in a detwinned one. The spinwave dispersion for this model is given by\bea\omega(h,k)=S\sqrt{A^2(h,k)-B^2(h,k)},\eea where\begin{widetext}
\bea A(h,k)&=&2[J_{1b}(\cos(\pi k)-1)+J_{1a}+2J_1+J_3(\cos(2\pi h)+\cos(2\pi k)-2)],\\\nonumber B(h,k)&=&2(J_{1a}\cos(\pi h)+2J_2\cos(\pi h)\cos(\pi k)).\eea \end{widetext}In the model, the dispersion along $[H,0]$-direction sees the maximum at $(1/2,0)$, while along $[1,K]$-direction, the maximum is at $(1,q)$ where $q\sim1/2$ is in fact parameter dependent. But one should not only compare $\omega(1/2,0)$ and $\omega(q,0)$ to find which direction reaches a higher top, because there is twinning. Once we have twinning into play, we also need to compare the band near $(0,1)$. Mark that in discussing this region, $H$ and $K$ directions should be interchanged when compared with the experimental frame. Starting from $(0,1)$, the dispersion reaches maximum along $K$-direction at $(0,1/2)$ and along $H$ direction at $(1/2,1)$. Therefore, in order to see what we see in the experiment, we should have \bea \max[\omega(1/2,0),\omega(0,1/2)]<\max[\omega(1,q),\omega(1/2,1)]\label{eq:inequality}.\eea

Now we make a statement and prove it: $J_{1b}<0$ is sufficient and necessary for the Eq.(\ref{eq:inequality}) to hold. First we prove the sufficiency.\bea\omega^2(1/2,1)-\omega^2(1/2,0)=16J_{1b}(-J_{1a}+J_{1b}-2J_2+2J_3).\eea From this we know if $J_{1b}<0$, we have $\omega(1/2,1)>\omega(1/2,0)$. On the other hand,\begin{widetext}\bea\omega^2(1/2,1)-\omega^2(0,1/2)=4(J_{1a}^2-2J_{1a}J_{1b}+J_{1b}(3J_{1b}-4J_2+4J_3)).\eea From this we know if\bea J_{1b}<\frac{1}{3}(J_{1a}+2J_2-2J_{3}-\sqrt{(J_{1a}+2J_2-2J_{3})^2-3J_{1a}^2},\eea $\omega(1/2,1)>\omega(0,1/2)$. \end{widetext}But of course\bea\frac{1}{3}(J_{1a}+2J_2-2J_{3}-\sqrt{(J_{1a}+2J_2-2J_{3})^2-3J_{1a}^2}>0,\eea therefore $J_{1b}<0$ is sufficient to make the highest energy along $[1,K]$ direction higher than $[H,0]$ direction.

Then we prove the necessity. It is a proof by contradiction. Suppose $J_{1b}>0$, then from above we know that $\omega(1/2,1)\le\omega(1/2,0)$. Also notice that when $J_{1b}>0$,\begin{widetext} \bea\omega^2(1,q_{y})&=&4(J_{1a}+2J_2+J_{1b}(\cos(\pi q)-1)-2J_3\sin^2(\pi q))^2-4(J_{1a}+2J_2\cos(\pi q))^2\\
\nonumber&<&4(J_{1a}+2J_2-2J_3)^2-4(J_{1a}-2J_2)^2\\
\nonumber&=&16(J_{1a}-J_3)(2J_2-J_3)\\
\nonumber&\le&4(J_{1a}+2J_2-2J_3)^2\\
\nonumber&=&\omega^2(1/2,0).\eea \end{widetext}Therefore\bea \max[\omega(1/2,0),\omega(0,1/2)]\ge\max[\omega(1,q),\omega(1/2,1)].\eea Therefore we have proved if $J_{1b}>0$ then the highest energy along $[1,K]$ direction is lower than the highest energy along $[H,0]$ direction, i.e., $J_{1b}<0$ is necessary to recover the feature observed in experimental data.

\begin{figure}[h]
\includegraphics[scale=.3]{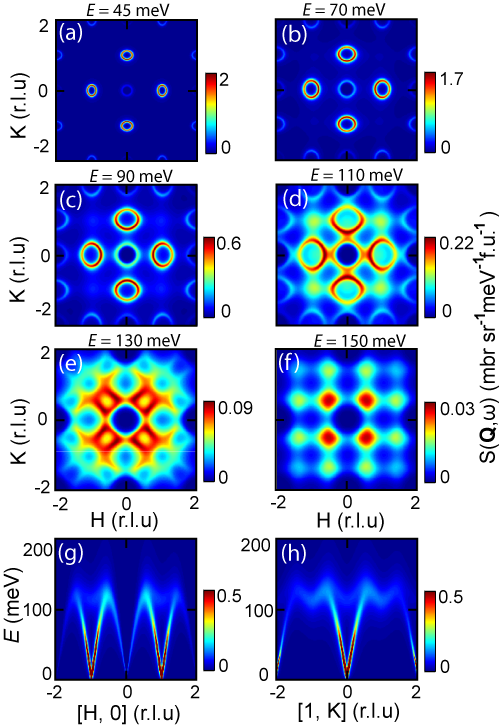}
\renewcommand{\figurename}{SI}
\caption{(color online). Constant-energy images of the scattering in the
$[H,K]$ zone as a function of increasing energy for Li$_{0.94}$FeAs at energy
transfers of (a) $45$ meV; (b) 70 meV;
(c) 90 mV; (d) 110 meV; (e) 130 meV; (f) 150
meV; (g,h) The expected spin wave dispersion along the $[H,0]$ and $[1,K]$ directions
for LiFeAs. The exchange couplings used to obtain these imagines are
$SJ_{1a}=23$ meV, $SJ_{1b}=-4$ meV, $SJ_{2}=13$ meV, $SJ_3=-5$ meV, and
anisotropy factor $SJ_s=0.08$ meV.  Spin wave damping is assumed to be
$\Gamma=0.15 E$ throughout the supplementary information.
}
\end{figure}

\begin{figure}[h]
\includegraphics[scale=.3]{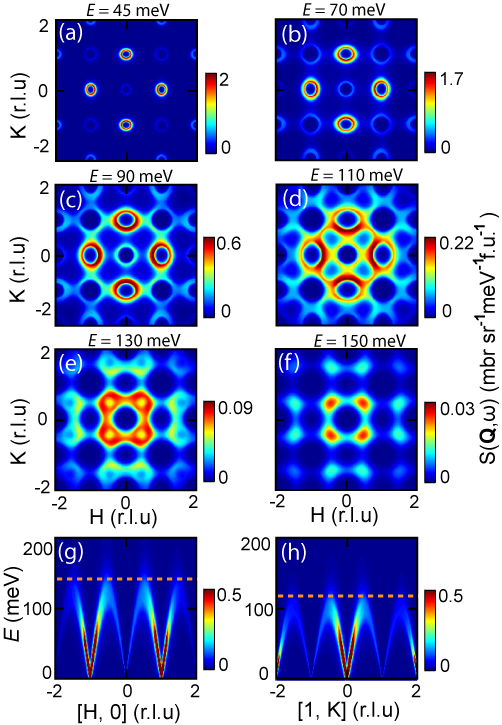}
\renewcommand{\figurename}{SI}
\caption{(color online). Constant-energy images of the scattering in the
$[H,K]$ zone as a function of increasing energy for Li$_{0.94}$FeAs at energy
transfers of (a) $45$ meV; (b) 70 meV;
(c) 90 mV; (d) 110 meV; (e) 130 meV; (f) 150
meV; (g,h) The expected spin wave dispersion along the $[H,0]$ and $[1,K]$ directions
for LiFeAs. The exchange couplings used to obtain these imagines are
$SJ_{1a}=10$ meV, $SJ_{1b}=10$ meV, $SJ_{2}=20$ meV, $SJ_3=-8$ meV, and
anisotropy factor $SJ_s=0.08$ meV.
}
\end{figure}

\begin{figure}[h]
\includegraphics[scale=.3]{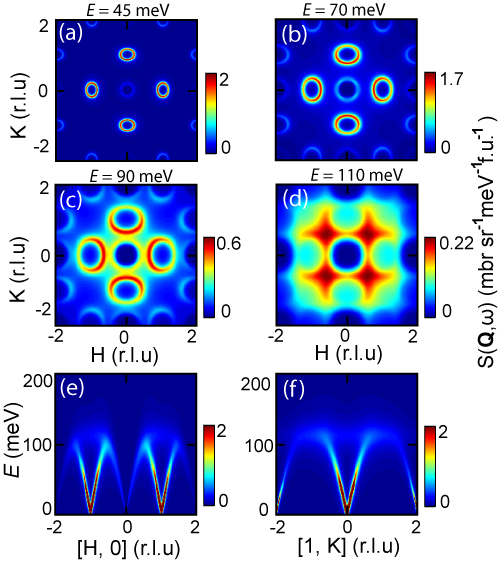}
\renewcommand{\figurename}{SI}
\caption{(color online). Constant-energy images of the scattering in the
$[H,K]$ zone as a function of increasing energy for Li$_{0.94}$FeAs at energy
transfers of (a) $45$ meV; (b) 70 meV;
(c) 90 mV; (d) 110 meV; (e,f) The expected spin wave dispersion along the $[H,0]$ and $[1,K]$ directions
for LiFeAs. The exchange couplings used to obtain these imagines are
$SJ_{1a}=23$ meV, $SJ_{1b}=-4$ meV, $SJ_{2}=13$ meV, $SJ_3=0$ meV, and
anisotropy factor $SJ_s=0.08$ meV.
}
\end{figure}

\begin{figure}[h]
\includegraphics[scale=.3]{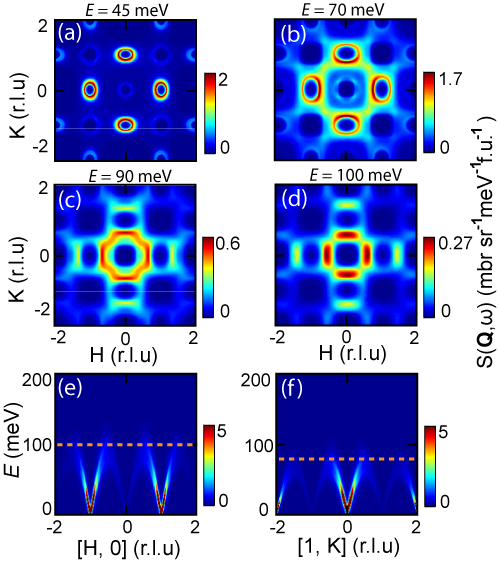}
\renewcommand{\figurename}{SI}
\caption{(color online). Constant-energy images of the scattering in the
$[H,K]$ zone as a function of increasing energy for Li$_{0.94}$FeAs at energy
transfers of (a) $45$ meV; (b) 70 meV;
(c) 90 mV; (d) 100 meV; (e,f) The expected spin wave dispersion along the $[H,0]$ and $[1,K]$ directions
for LiFeAs. The exchange couplings used to obtain these imagines are
$SJ_{1a}=10$ meV, $SJ_{1b}=10$ meV, $SJ_{2}=20$ meV, $SJ_3=0$ meV, and
anisotropy factor $SJ_s=0.08$ meV.
}
\end{figure}

\end{document}